\documentclass{WileyMSP-template}
\usepackage{amsmath}
\usepackage{amssymb}
\usepackage{braket}
\usepackage[super,square,compress,sort]{natbib}

\usepackage{setspace}
\usepackage{lineno}

\begin{document}


\title{Topological monopole’s gauge field induced anomalous Hall effect in artificial honeycomb lattice}

\maketitle

\author{Jiasen Guo}
\author{Vitalii Dugaev}
\author{Arthur Ernst}
\author{George Yumnam}
\author{Pousali Ghosh}
\author{Deepak Kumar Singh$^\ast$}

\begin{affiliations}
J. Guo, G. Yumnam, P. Ghosh, Prof. D. K. Singh\\
Department of Physics and Astronomy\\
University of Missouri, Columbia \\
Columbia, Missouri 65211, USA \\
E-mail: singhdk@missouri.edu\\
\medskip
Prof. V. Dugaev\\
Department of Physics and Medical Engineering\\
Rzeszów University of Technology\\
Rzeszów, 35-959, Poland\\
\medskip 
Prof. A. Ernst\\
Institut f\"{u}r Theoretisce Physik\\
Johannes Kepler Universit\"{a}t\\
Linz, 4040, Austria\\
Max-Planck-Institut f\"{u}r Mikrostrukturphysik\\
Halle, 06120, Germany\\
\end{affiliations}

\section*{Key Points}
\begin{enumerate}
    \item{Demonstration of gauge field flux due to vortex magnetism.}
    \item{Anomalous Hall effect in magnetic honeycomb lattice.}
    \item{Topological monopole induced magnetic flux.}
\end{enumerate}
\clearpage

\justify
\section*{Abstract}
\begin{abstract}
Vortex magnetic structure in artificial honeycomb lattice provides a unique platform to explore emergent properties due to the additional Berry phase curvature imparted by chiral magnetization to circulating electrons via direct interaction. We argue that while the perpendicularly-aligned magnetic component leads to the quantized flux of monopole at the center of the Berry sphere, the in-plane vortex circulation of magnetization gives rise to unexpected non-trivial topological Berry phase due to the gauge field transformation. The unprecedented effect signifies the importance of vector potential in multiply-connected geometrical systems. Experimental confirmations to proposed hypotheses are obtained from Hall resistance measurements on permalloy honeycomb lattice. Investigation of the topological gauge transformation due to the in-plane chirality reveals anomalous quasi-oscillatory behavior in Hall resistance $R_{xy}$ as function of perpendicular field. The oscillatory nature of $R_{xy}$ is owed to the fluctuation in equilibrium current as a function of Fermi wave-vector $k_F$, envisaged under the proposed new formulation in this article. Our synergistic approach suggests that artificially tunable nanostructured material provides new vista to the exploration of topological phenomena of strong fundamental importance.
\end{abstract}

\section*{Introduction}
Electrons traversing a metallic loop can result in the acquisition of a net geometric phase to its wavefunction when a parameter, such as the potential energy, is slowly varied.\cite{Xiao,McDonald,Liu,Son} The geometric phase acquired by the particle is equal to the flux of a new field called the Berry curvature through the surface defined by this loop.\cite{Son,Haldane,Weng} In magnetic materials with chiral moment arrangement, electron’s scattering from magnetic moment renders additional Berry curvature due to the Dirac’s magnetic monopole flux through the surface defined by the closed path.\cite{Chen,Zhang,Felser} Unlike the Aharonov-Bohm effect, which provides theoretical foundation for the observation of monopoles in real space, the Berry curvature strictly occurs in reciprocal space.\cite{McDonald,Rosch} The elegant formulation of Berry curvature arising due to the divergence of flux from the Dirac’s monopole led to broad synergistic exploration of the fundamental entity in reciprocal space in solid state materials. It includes the investigation of simple perovskite SrRuO$_3$ to unconventional magnets, such as chiral hedgehog, skyrmions and spin ice (although the underlying physical mechanism behind Dirac’s magnetic charge in spin ice is arguably different).\cite{Zhang,Tokura,Volovik,Milde,Nayak,Bramwell}\\

The Berry curvature can also have its origin in the topological spin chirality, described by the scalar triple product.\cite{Nayak,Nagaosa} For instance, the observation of anomalous Hall effect (AHE) in the molybdate pyrochlore revealed the topological nature of Hall signal due to the Berry curvature.\cite{Nagaosa,Tokura2} The AHE effect with topological characteristic was also envisaged to be manifested in a two-dimensional array of nanomagnetic cylinders.\cite{Vitalii} The magnetization direction is aligned along the cylinder’s length, perpendicular to the thin metallic film where Hall conductivity is measured. Now, imagining a ring-shaped magnetic system, consisting of both an in-plane vortex magnetic configuration and a canted chiral moment structure perpendicular to the plane of the ring, can in principle lead to the detection of both the AHE associated to the net Berry curvature and the magnetic monopole’s quantization directly. While the AHE is due to the vortex in-plane magnetization pattern, the magnetic monopole’s quantization can be associated to the canted moment configuration.\cite{Vitalii2} The physical mechanism is schematically described in Figure 1a and b. Figure 1a describes the closed contour C along the ring in the $x$-$y$ plane, a mapping of which on the Berry sphere is shown in Figure 1b. In the figure, magnetization orientation has two components – along the ring (in-plane) and along the $z$-axis (out-of-plane). Such unique scenario can be realized in artificially created two-dimensional magnetic (permalloy) honeycomb lattice,\cite{Heyderman} made of nanoscopic connected elements ($\sim$ 12 nm in length).\cite{Summers,Artur} The strong shape anisotropy in permalloy element ensures that the system maintains an in-plane component of magnetization even in the large perpendicular field application (it would require a field larger than the magnitude of 1/2 $\mu V^2$, $\mu$ $\sim$ 100,000 erg/cm$^2$ for permalloy, to completely align the moment to the perpendicular field direction).\cite{Heyderman,Blundell}

\section*{Methods}
\subsection*{Sample Fabrication} 
We create Permalloy (Py) and Permalloy-Platinum (Py-Pt) honeycomb lattice samples using diblock template method, which results in large throughput sample with ultra-small connecting elements of $\sim$ 12 nm in length. Details about the synthesis of diblock template can be found somewhere else.\cite{Russell} The diblock template is fabricated by spin coating copolymer PS-b-P4VP on silicon substrate, followed by solvent vapor annealing. The resulting diblock template resembles a honeycomb lattice with a typical element size of 12 nm (length) and 5 nm (width), as is shown in Figure 1c. This topographical property was exploited to create metallic honeycomb lattice by depositing permalloy, Ni$_{0.81}$Fe$_{0.19}$, in near parallel configuration in an electron-beam evaporator. The substrate was rotated at a moderate constant speed about its axis during the deposition process to create uniformity in the film thickness. In the case of Py-Pt sample, thin layer of Pt is deposited on top of Py layer without breaking the vacuum, thus ensuring the clean contact between two layers. 
\subsection*{Electrical Measurement}
Hall probe measurements are performed on a 2$\times$2 mm$^2$ sample using the standard contact configuration, as prescribed by National Institute of Standards and Technology,\cite{NIST} see inset in Figure 1. Electrical measurements were performed in a cryogen-free 9 T magnet with a base temperature of $\sim$ 5 K, using a set of current source and nanovoltmeter from Keithley.

\section*{Results}
The circulation of gauge field $\boldsymbol{\mathcal{A}}$ along the contour C is the topological Berry phase, generated by the uniform magnetization. Typically, the Berry phase is related to the mapping $\mathbf{r} \rightarrow \mathbf{n(r)}$ of contour C to the Berry sphere (see Figure 1b), with the magnetic monopole in the center of this sphere.\cite{Xiao,He,McDonald} Consequently, the flux $\gamma_C$ of $\boldsymbol{\mathcal{B}}$ field through the surface, limited by the contour C is given by
\begin{equation}
    \gamma_c=\oint{\boldsymbol{\mathcal{A}}\cdot d\mathbf{l}}=\int{\boldsymbol{\mathcal{B}}\cdot d\mathbf{S}}
\end{equation}
where $\boldsymbol{\mathcal{B}}(\mathbf{r})=\boldsymbol{\nabla}\times \boldsymbol{\mathcal{A}}(\mathbf{r})$ is the Berry curvature. The gauge field $\boldsymbol{\mathcal{A}}$ is a matrix in spin space, defined by $\boldsymbol{\mathcal{A}}(\mathbf{r})=iU(\boldsymbol{\nabla} U^{-1})$ where $U(\mathbf{r})$ is a unitary transformation such that $U(\boldsymbol{\sigma}\cdot \mathbf{n})U^{-1}=\boldsymbol{\sigma}\cdot \mathbf{n_0}$, $\mathbf{n_0}$ is a constant vector (see Supplementary Materials for detail). In the case of magnetization along the ring with a constant value of $n_z$ (see Figure 1b), the matrix of unitary transformation is $U=e^{(i\alpha\sigma_z)}$ where $\alpha (\mathbf{r})$ is the angle around the circular contour. Now, in the presence of external magnetic field $\mathbf{B_{ex}}$, the quantity $\boldsymbol{\mathcal{A}}$ undergoes a gauge transformation of 
\begin{equation}
    \boldsymbol{\mathcal{A}}\rightarrow \boldsymbol{\mathcal{A}}+\frac{eA_l}{\hbar c},
\end{equation}
where $A_l=B_{ex}R/2$ is the longitudinal component (along the ring) of the electromagnetic field vector potential $\mathbf{A}$. Correspondingly, the net Berry phase is replaced by
\begin{equation}
    \gamma_C\rightarrow \gamma_C+\frac{e\Phi}{\hbar c}=2\pi\left ( 1+\frac{\Phi}{\Phi_0} \right ),
\end{equation}
where magnetic flux $\Phi$ due to the external magnetic field $\mathbf{B_{ex}}$ is given by $\pi R^2B_{ex}$ and $\Phi_0$ is the quantized flux of $hc/e$. The change in Berry phase in applied field basically infers a quantized change in the topological magnetic charge at the center of the Berry sphere. A mathematical confirmation to this effect is obtained by calculating the energy of electron in gauge transformed vector potential $\boldsymbol{\mathcal{A}}$. As described in Supplementary Materials, the energy of electron in a quantum level $s$ depends on the angle $\theta$ between magnetization vector $\mathbf{M}$ and the $z$-axis via the following expression:
\begin{equation}
    \frac{\varepsilon_s}{\varepsilon_0}=\frac{\tilde{s}^2+1}{2}\pm(\tilde{s}^2-2\tilde{s}\xi t+\xi^2)^{1/2},
\end{equation}
where 
\begin{equation*}
\varepsilon_0=\frac{\hbar^2}{mR^2},\qquad \xi=\frac{gM_0}{\varepsilon_0},\qquad t=\cos(\theta). 
\end{equation*}
The quantum state $\tilde{s}$ is defined as $\tilde{s}=s-\Phi/\Phi_0$. The dependence of $\varepsilon_s$ with $s = 1$ on $t$ for different values of field is presented in Figure 2a. In Figure 2a, we see that when the magnetic field is small, the electron energy has a minimum at $t = 0$ (for in-plane magnetization). But with the increasing of the field, after certain critical value $\Phi_c$ of the flux, there appear an energy gain related to electron system, which makes it favorable for magnetization to cant out of in-plane configuration. Besides, the jump associated to the critical flux $\Phi_c$ imposes a relation between magnetic field $B_c$ and the cell size $S$. The obtained results can be interpreted in terms of the Berry phase of electron moving along the contour C within the ring. Indeed, the wave function of electron moving adiabatically in a non-homogeneous magnetization field $\mathbf{M(r)}$ acquires a net Berry phase, given by, $\gamma_C^g=\int_C\mathcal{A}_l\cdot dl$, where $\mathcal{A}_l$ is the longitudinal component of gauge potential. For the closed contour (ring) we find, $\gamma_C^g=\oint \mathcal{A}_l\cdot dl=2\pi$, which is one-half of the full surface of Berry sphere (the mapping space of vector field $\mathbf{n(r)}$). This corresponds to the flux of topological (gauge) field $\boldsymbol{\mathcal{B}}=\boldsymbol{\nabla}\times \boldsymbol{\mathcal{A}}$ created by the monopole with topological charge $e_m = 1$ in the center of the Berry sphere. The magnetic field $\mathbf{B}$ generates an additional (Aharonov-Bohm) phase $\gamma_C^m=\oint A_l\cdot dl$ across the same closed contour C such that the total flux $\gamma_C=\gamma_C^g+\gamma_C^m$. This can be viewed as a variation of monopole charge, which changes the total flux through one-half of the Berry sphere. 

As discussed in the previous paragraph, the gauge field arising due to the vortex structure of magnetic configuration can lead to the topological effect of net Berry phase accumulation. Experimental confirmation to this effect can be obtained from the Hall effect measurements. We have performed Hall measurements on permalloy (Py) honeycomb samples (see Sample Fabrication Section for details about the nanofabrication procedure). Hall effect magnetoresistance measurements on a $2\times2$ mm$^2$ square size permalloy honeycomb sample reveal symmetric quasi-oscillatory responses in $R_{xy}$ as a function of magnetic field at low temperature $T$ = 5 K, see Figure 3a. The Hall resistance $R_{xy}$ increases with field, peaking around $H$ $\sim$ 1.5 T before decreasing again. The observation is in stark contrast to the linear $R_{xx}$, as shown in Figure 3b, which manifests surprising negative magnetoresistance as a function of field. We notice a sharp decrement in $R_{xx}$ below $H$ $\sim$ 0.5 T followed by the gradual decline as magnetic field increases. While a clear explanation to this effect is lacking, it is most likely arising due to the weak localization of electrons, interacting with magnetization $\mathbf{m}$ via $\boldsymbol{\sigma}\cdot \mathbf{m}$, in the honeycomb element.\cite{McCann,Vitalii3} Note that localization effects are strongly enhanced in nanostructures if the phase relaxation length $l_\phi$ is larger than the nanoelement thickness $d$. In magnetic honeycomb lattice, $d$ $\sim$ 5 nm. The magnetic field application suppresses the localization corrections, thus leading to the negative magnetoresistance. Unlike the field dependence, $R_{xy}$ and $R_{xx}$ don’t exhibit any unusual behavior as a function of temperature. As shown in Figure 3c and d, the system manifests semiconducting characteristic in both zero and applied field. 

The quasi-oscillatory behavior in $R_{xy}$ becomes conspicuously prominent in permalloy sample coated with a thin layer ($\sim$ 2 nm) of platinum (Py-Pt), see Figure 4a. The pronounced quasi-oscillatory characteristic of Hall resistance in Py-Pt sample can be attributed to the spin-orbital (SO) coupling due to the Pt film (as discussed below).\cite{Huda,Vitalii3} However, the change in Hall resistance is modest compared to that found in only Py honeycomb sample.  In the Py-Pt sample, the Hall resistance does not seem to be symmetric in field. Rather, the peak in $R_{xy}$ occurs at $H$ $\sim$ 0.5 T on the positive side and at $H$ $\sim$ $-3.8$ T on the negative side. Unlike the Py sample, the linear magnetoresistance in Py-Pt sample exhibits positive enhancement as a function of field, following a negative tendency at $H$ $<$ 0.5 T, see Figure 4b. The experimental observation is in accord with the conventional understanding of Py thin film, which is known to exhibit positive magnetoresistance. Pt coating seems to alter the electrical characteristic of honeycomb lattice sample at low temperature. Unlike the Py sample, manifesting semiconducting behavior, Py-Pt sample shows weakly metallic property as temperature reduces, see Figure 4c and d. The observation could be attributed to the weak localization effect, in addition to the SO coupling.\cite{Vitalii3} Due to the SO coupling, the system tends to reflect the metallic characteristic of Pt film. However, the strong semiconducting behavior of Py honeycomb lattice dominates at low temperature. 

\section*{Discussion}
There are two conceptual issues here: the oscillatory behavior in $R_{xy}$ and a negative linear magnetoresistance in field. The oscillatory behavior in Hall resistance and its possible topological origin can be understood from the first principle calculations. For this purpose, we consider the model Hamiltonian of
\begin{equation}
    \left [ -\frac{\hbar^2(\nabla_x^2+\nabla_y^2)}{2m}+g\boldsymbol{\sigma}\cdot \mathbf{m(r)}-\varepsilon  \right ]\psi(\mathbf{r}) = 0,
\end{equation}
where $\mathbf{m(r)}=\frac{\lambda_0}{r^2}(-y,x)=\frac{\lambda_0}{r^2}\mathbf{\hat{z}}\times \mathbf{r}$ represents local magnetization in the loop state, shown in Figure 1d. The solution to this equation is $\psi^T_n(\alpha)=(c_1e^{in\alpha}, c_2e^{i(n+1)\alpha})$. Correspondingly, the Berry phase on the contour along the ring can be written as
\begin{equation}
    \phi_n=\int_0^{2\pi}{ A_n(\alpha)d\alpha},
\end{equation}
where $A_n(\alpha)$ is the gauge potential, given by $A_n(\alpha)=-i\psi_n^\dagger\nabla_\alpha\psi_n$. The Berry phase is used to calculate the energy spectrum, equilibrium current and off-diagonal conductivity (Hall resistance) as functions of chemical potential and the parameter $k_Fa_0$. Plots of $\sigma_{xy}$ vs. $k_Fa_0$ and  $j_\alpha$ vs. $\mu$ are shown in Figure 5 (see Supporting Information for detail). Now, the Hall resistance $R_{xy}$ is dependent on $\sigma_{xy}$ via the following relation
\begin{equation}
   R_{xy}\simeq -\frac{\sigma_{yx}}{\sigma^2_{xx}}\simeq -\sigma_{yx}R_0^2\left (  1+\frac{2\delta R(H)}{R_0}  \right ), 
\end{equation}
where 
\begin{equation}
\sigma_{xx}=\frac{1}{R_{xx}}\simeq \frac{1}{R_0}-\frac{\delta R(H)}{R_0^2} 
\end{equation}
with $R_{xx}=R_0+\delta R(H)$. In the experiment, the observed absolute $\delta R(H) \ll R_0$, thus, the oscillatory behavior in $\sigma_{xy}$ due to the vortex configuration of in-plane magnetization, as shown in Figure 5a, is reflected in the oscillatory Hall resistance $R_{xy}$, albeit weakly. Also, it is worth noting that magnetic field application upends the chemical potential, hence the Fermi parameter $k_F$ and the energy spectrum $\varepsilon_k$. Cumulatively, it can be concluded that the vortex configuration of in-plane magnetization, as found in magnetic honeycomb lattice at low temperature, manifests topological characteristic of Berry phase accumulation. 

\section*{Conclusion}
Finally, we summarize the main results of the paper. The synergistic study presented here not only elucidates the occurrence of magnetic charge induced gauge transformed flux through a ring-shaped contour of perpendicular moment in honeycomb lattice but also reveals a new mechanism behind the topological nature of chiral vortex circulation due to in-plane magnetization. It should be noted that the proposed effect of the gauge field, related to inhomogeneous magnetization in the ring, cannot be reduced to the gauge-field-induced AHE. The latter mechanism would be similar to magnetic-field-induced classical Hall effect. As we demonstrated, the gauge field in the nanoring is substantially affecting the wave functions and the energy spectrum of electrons. As a result, we come to a quantum AHE in the magnetic nanostructure. This is reminiscent of the quantum Hall effect, in which the field-induced transverse current is transferred by the quantum excitations of 2D electron system (not by free electrons) in the magnetic field. The vortex magnetic configuration causes an additional flux to the one due to magnetic charge at the center of the corresponding three-dimensional Berry sphere. The net gain in Berry phase of cycling electrons due to the in-plane magnetization is highly surprising. The non-trivial effect suggests strong implication of the vector potential $\mathbf{A}$ in electrodynamics problems. Typically, the AHE due to the Berry phase, as found in pyrochlore compounds,\cite{Nagaosa} is ascribed to the chirality due to the scalar triple product of magnetic spins. In the case of in-plane vortex moment configuration, the scalar triple product is zero. However, as we have seen, the net Berry phase is non-zero, which gives rise to the AHE signal. It clearly suggests that a new mechanism is at the play. We believe this effect can be realized in other two-dimensional multiply connected systems as well.

\medskip
\noindent\textbf{Supporting Information} \par
\noindent Supporting Information is available from the Wiley Online Library or from the author.

\medskip
\noindent\textbf{Acknowledgements} \par
\noindent DKS thankfully acknowledges the support by US Department of Energy, Office of Science, Office of Basic Energy Sciences under the grant no. DE-SC0014461. VKD thankfully acknowledges the support by the National Science Center in Poland, research project No. DEC-2017/27/B/ST3/02881.

\medskip
\noindent\textbf{Author Contributions} \par
\noindent DKS envisaged the research concept and supervised every aspect of research. JG and PG fabricated the magnetic honeycomb samples. JG and GY performed the Hall effect measurements. VKD and AE performed the first principle calculations. DKS prepared the manuscript where everyone contributed.

\medskip
\noindent\textbf{Data Availability Statement}\par
\noindent The data that supports the findings of the study are available from the corresponding author upon reasonable request.

\medskip
\noindent\textbf{Conflict of Interest}\par
\noindent The authors declare no potential conflicts of interest.

\medskip
\noindent\textbf{ORCID}\par
\noindent Jiasen Guo, https://orcid.org/0000-0001-7708-1476\par
\noindent Vitalii Dugaev, https://orcid.org/0000-0002-0999-4051 \par
\noindent Arthur Ernst, https://orcid.org/0000-0003-4005-6781 \par
\noindent George Yumnam, https://orcid.org/0000-0001-9462-7434\par
\noindent Pousali Ghosh, https://orcid.org/0000-0002-3393-9941 \par


\clearpage

\begin{figure}
    \centering
    \includegraphics[width=\linewidth]{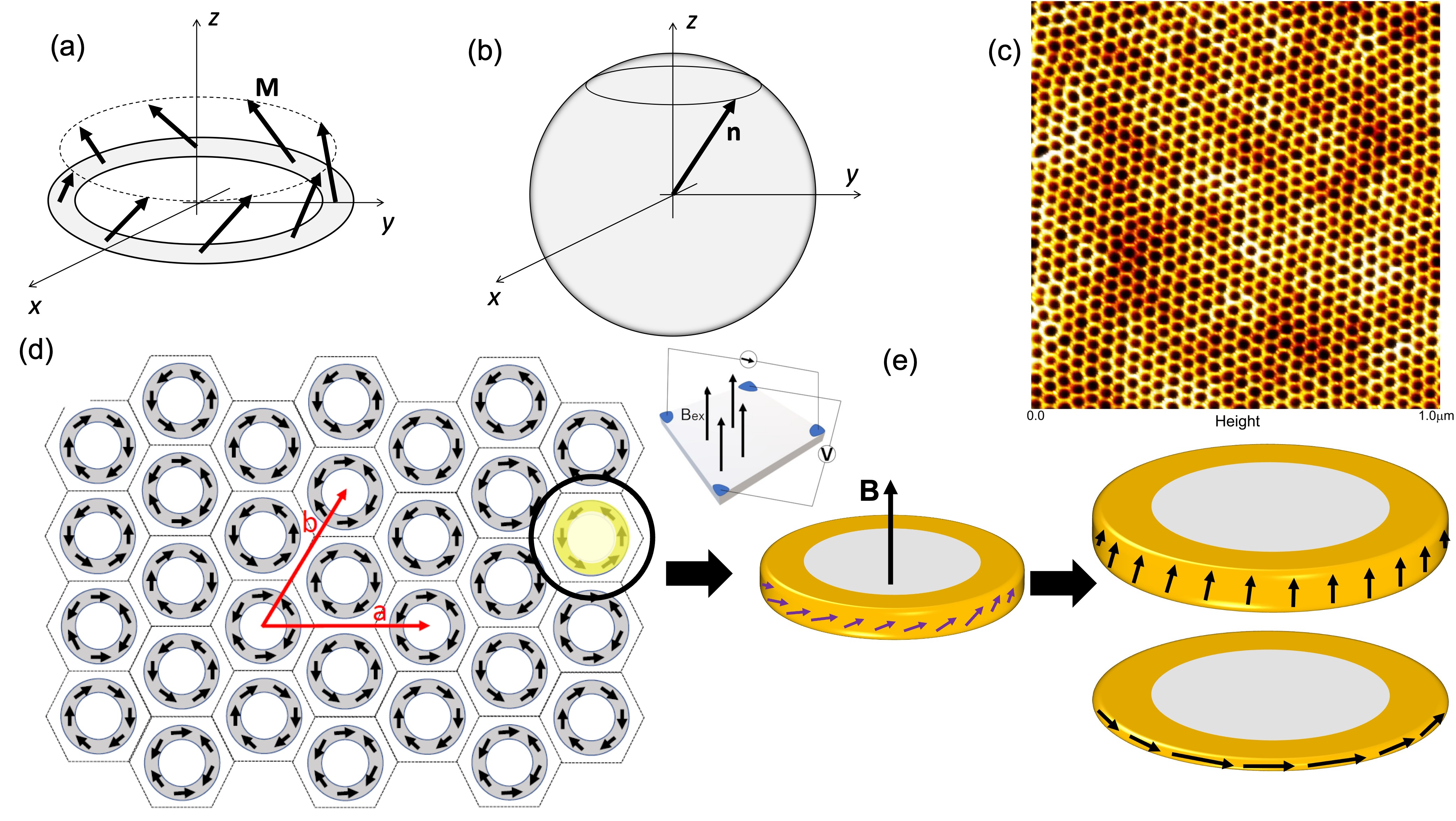}
    \caption{\textbf{Berry phase due to magnetization profile in a ring.} (a) Magnetization along the ring in magnetic field applied perpendicular to the plane of the system. Magnetization has both in-plane and perpendicular components. (b) Mapping of the contour due to perpendicular component across ring on the Berry sphere. (c) Atomic force micrograph of a nanoscopic honeycomb lattice. The typical size of permalloy element, used in this study, is 12 nm. (length) 5 nm (width). For the Py only sample, $\sim$ 6 nm Py are deposited; for the Py-Pt sample, $\sim$ 6 nm Py and 2 nm Pt are deposited. (d)  Permalloy honeycomb lattice is known to manifest spin solid state due to vortex magnetic loop at low temperature, schematically described here. (e) Perpendicular field application causes moments to cant out of the plane. However, strong shape anisotropy in permalloy competes against field-induced alignment of moment along $z$-axis. This unique scenario allows for the direct demonstrations of monopole’s induced gauge transformed flux due to Berry phase in vortex magnetic configuration, which causes anomalous Hall effect in permalloy honeycomb lattice. Inset shows the Van der Pauw configuration used for the Hall effect measurement.}
    \label{fig:Figure_1}
\end{figure}

\begin{figure}
    \centering
    \includegraphics[width=\linewidth]{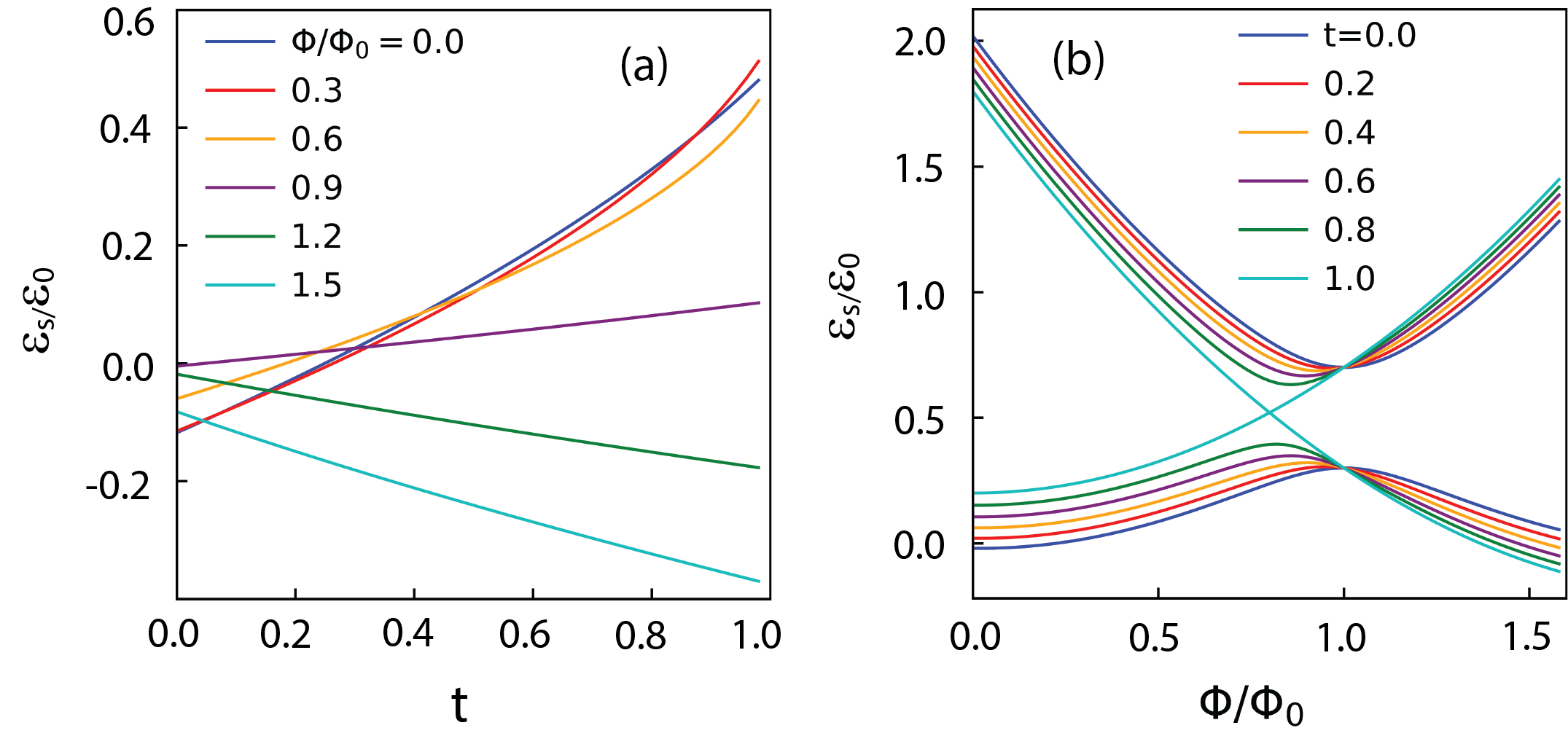}
    \caption{\textbf{Energetics behind monopole’s gauge transformed flux.} (a) Electron energy levels in the state with $s$ = 1 as a function of magnetization orientations for different magnetic fields. The parameter $\xi$ = 0.5. (b) Electron energy in the state with $s$ = 1 as a function of magnetic field for different magnetization orientations. The parameter $\xi$ = 0.2. }
    \label{fig:Figure_2}
\end{figure}

\begin{figure}
    \centering
    \includegraphics[width=\linewidth]{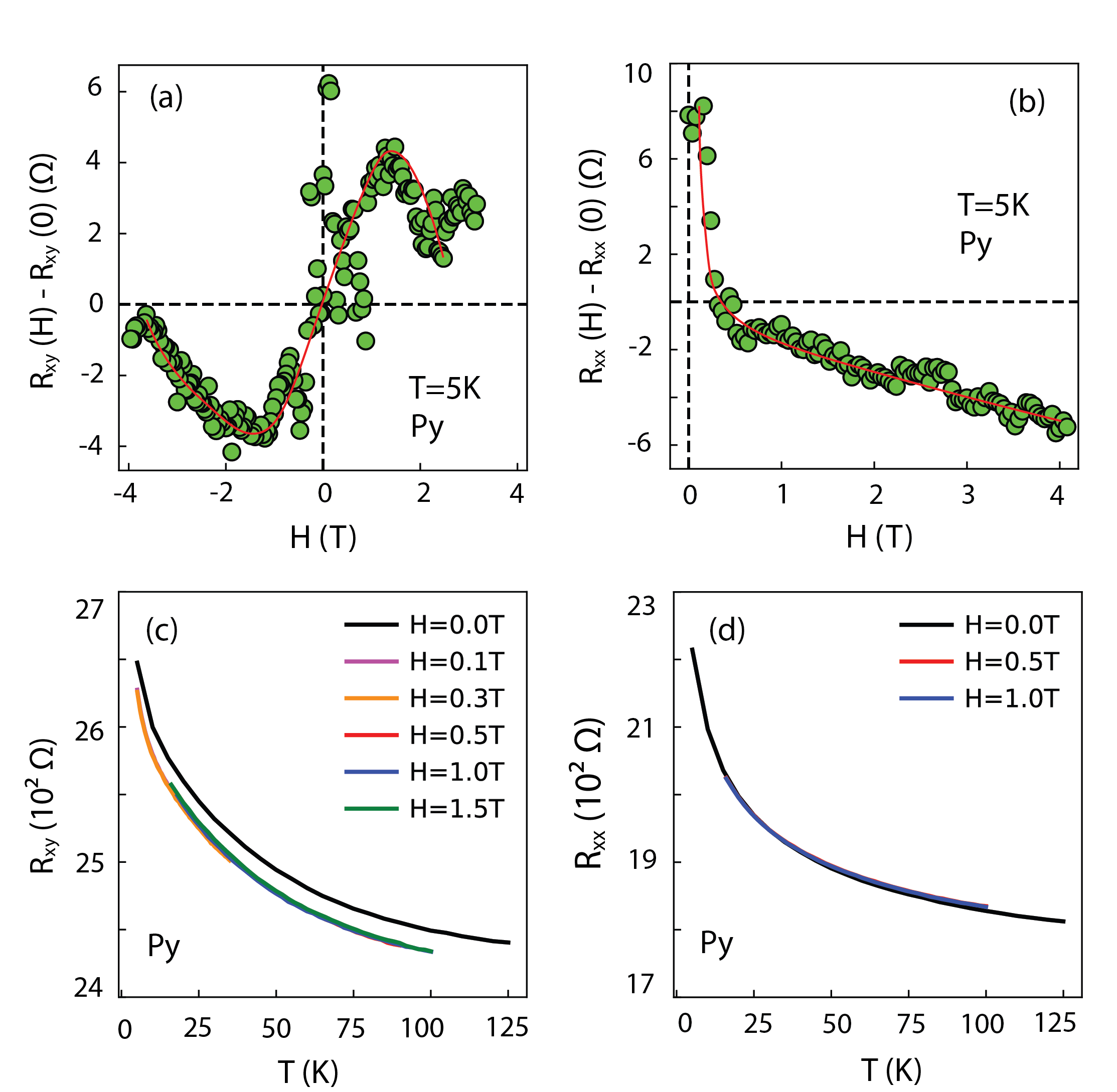}
    \caption{\textbf{Anomalous Hall effect due to topological characteristic of vortex loop of in-plane moment structure.} In a non-trivial depiction of vector potential’s role in generating Berry phase due to vortex magnetic profile, we show Hall resistance $R_{xy}$ data (Figure a) and linear resistance (Figure b) from permalloy (Py) honeycomb lattice at $T$ = 5 K. The red lines are for eye guidance. Oscillatory behavior is detected in Hall data in ($R_{xy}(H)$ – $R_{xy}$(0)) plot, while the linear resistance manifests negative MR as a function of field. Measurements of both $R_{xy}$ and $R_{xx}$ reveal semiconducting characteristic of Py honeycomb lattice system. }
    \label{fig:Figure_3}
\end{figure}

\begin{figure}
    \centering
    \includegraphics[width=\linewidth]{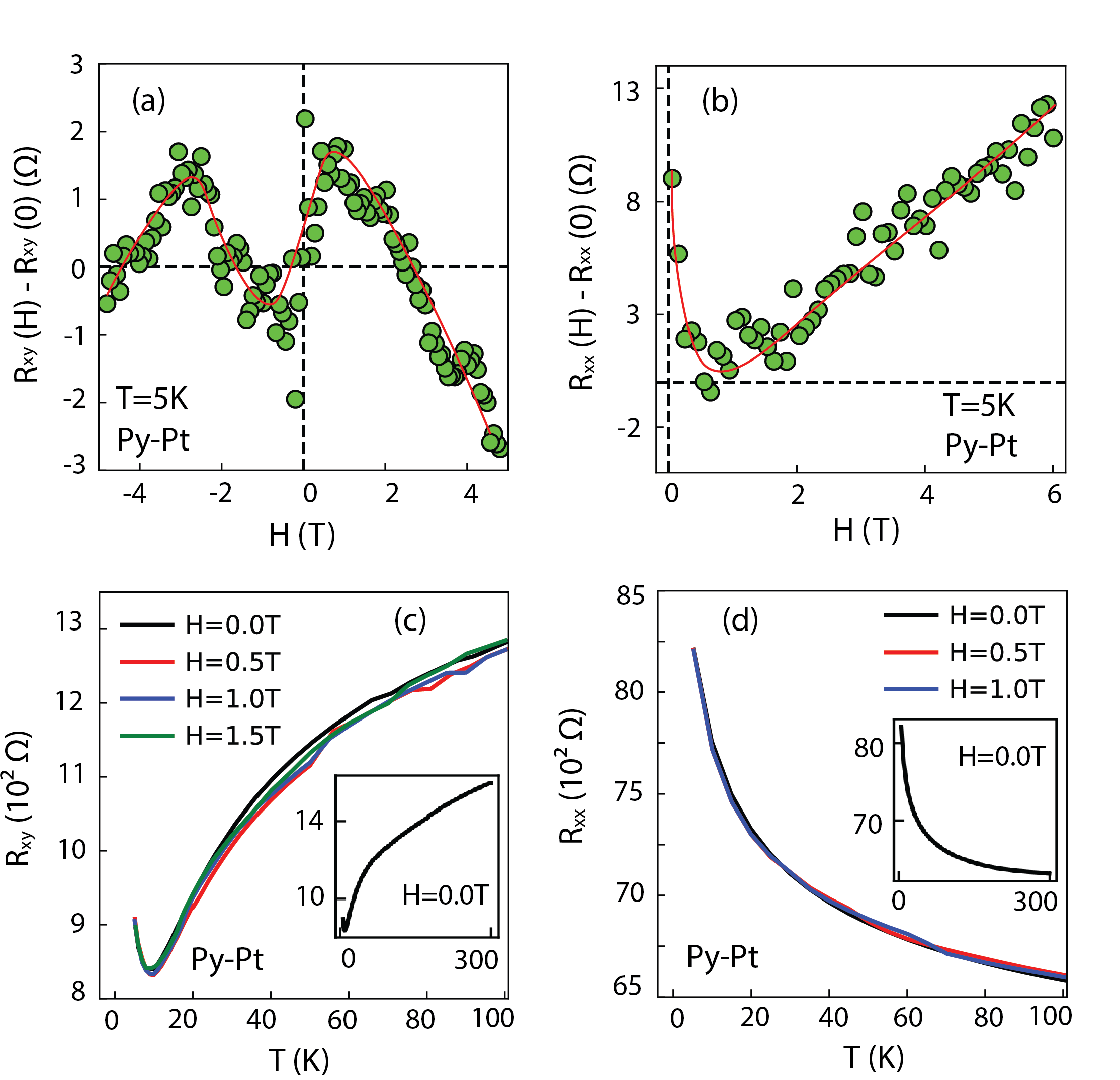}
    \caption{\textbf{Anomalous Hall effect in Py-Pt system.} Plots of Hall resistance $R_{xy}$ (Figure a) and linear resistance (Figure b) measurements on Py-Pt sample at $T$ = 5 K. Unlike Py sample, quasi-oscillation behavior in ($R_{xy}(H)$ – $R_{xy}$(0)) is more apparent in Py-Pt sample due to the spin-orbital coupling. Also, contrasting magnetoresistance tendency is observed in Py-Pt honeycomb in linear resistance measurements of $R_{xx}$ (Figure b). The red lines are for eye guidance. Additionally, Py-Pt sample exhibits weak metallic characteristic (Figure c) in $R_{xy}$ data as a function of temperature, which is different from the Py honeycomb sample. Figure d shows the semiconducting behavior in $R_{xx}$ vs. $T$ (K) plot, similar to Py sample.}
    \label{fig:Figure_4}
\end{figure}

\begin{figure}
    \centering
    \includegraphics[width=\linewidth]{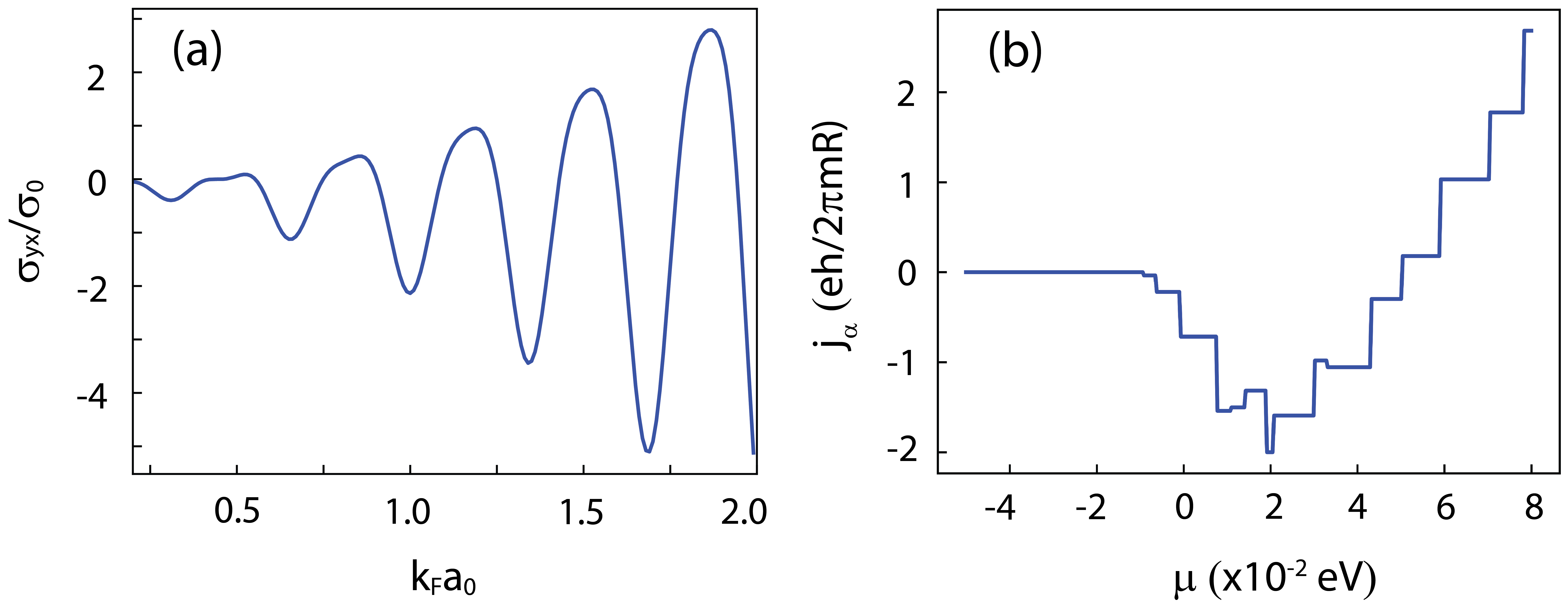}
    \caption{\textbf{Theoretical manifestation of oscillatory behavior in Hall conductivity.}  (a) First principal calculation reveals the oscillatory behavior in transverse conductivity as a function of $k_Fa_0$. (b) Equilibrium current in the magnetized ring as a function of chemical potential.}
    \label{fig:Figure_5}
\end{figure}

\end{document}



\title{Supplementary materials: Topological monopole's gauge field induced anomalous Hall effect in artificial honeycomb lattice}

\maketitle

\author{Jiasen Guo}
\author{Vitalii Dugaev}
\author{Arthur Ernst}
\author{George Yumnam}
\author{Pousali Ghosh}
\author{Deepak Kumar Singh$^\ast$}

\begin{affiliations}
J. Guo, G. Yumnam, P. Ghosh, Prof. D. K. Singh\\
Department of Physics and Astronomy\\
University of Missouri, Columbia \\
Columbia, Missouri 65211, USA \\
E-mail: singhdk@missouri.edu\\
\medskip
Prof. V. Dugaev\\
Department of Physics and Medical Engineering\\
Rzeszów University of Technology\\
Rzeszów, 35-959, Poland\\
\medskip 
Prof. A. Ernst\\
Institut f\"{u}r Theoretisce Physik\\
Johannes Kepler Universit\"{a}t\\
Linz, 4040, Austria\\
Max-Planck-Institut f\"{u}r Mikrostrukturphysik\\
Halle, 06120, Germany\\
\end{affiliations}
\clearpage

\justify
\section{Berry phase due to the magnetization profile in the ring}

Herewith we propose a possible mechanism of experimentally observed
magnetization jumps with the variation of the applied magnetic field.
This effect is related to the Berry phase due to inhomogeneity of the
in-plane magnetization in the rings. The idea is to account for the
energy of an electron system under the gauge potential related to the
magnetization and the vector potential of the external field
${\bf B}$.  Indeed, the total energy $E$ of the system includes the
magnetic energy $E_m$ and the energy of electron system $E_{el}$. Both
of them are depending on the magnetic field ${\bf B}$ and the
magnetization distribution ${\bf M}({\bf r})$. At low temperatures,
one realizes a ``saddle-point" magnetization profile
${\bf M}_{sp}({\bf r})$ corresponding to the minimum of the total
energy $E$ at a given ${\bf B}$.  A nonlinear dependence of
${\bf M}_{sp}({\bf r})$ on the external field ${\bf B}$ can be the
reason of the magnetization jumps.

To make it clear we consider a model with electrons on the magnetic
ring. In the absence of an external field, the
magnetization is in-plane along the ring due to magnetic anisotropy,
so that $M_z=0$. When we apply the field ${\bf B}$ along axis $z$, the
magnetization can acquire a $z$-component $M_z$ (see Figure 1a). To estimate the energy of the electron system we consider electrons on the
quantum ring. The corresponding Hamiltonian can be presented as  
\begin{eqnarray}
\label{1}
H=\int d^2{\bf r}\; \psi ^\dag \Big[ -\frac{\hbar ^2}{2m}\Big( \boldsymbol{\nabla} -\frac{ie{\bf A}}{\hbar c}\Big)
+W({\bf r})-gM_0\, \bsig \cdot {\bf n}({\bf r})\Big] \, \psi ,
\end{eqnarray} 
where $\psi ^\dag ({\bf r})$ and $\psi ({\bf r})$ are the spinor
creation and annihilation operators, ${\bf A}({\bf r})$ is the
electromagnetic vector potential, ${\bf n}({\bf r})$ is the unit vector of magnetization, ${\bf M}({\bf r})=M_0{\bf n}({\bf r})$, and $g$ is the coupling constant. The potential $W({\bf r})$ bounds electrons to the ring. One can use a local unitary transformation $U({\bf r})$, $\psi \to U\psi $ such that  
\begin{eqnarray}
\label{2}
U\, (\bsig \cdot {\bf n})\, U^{-1}=\bsig \cdot {\bf n}_0\, .
\end{eqnarray}
where ${\bf n}_0$ is a constant unit vector, and we choose ${\bf n}_0=(n_{0x},\, 0,\, n_{0z}$).
Then after $U$-transformation we get
\begin{eqnarray}
\label{3}
H=\int d^2{\bf r}\; \psi ^\dag \Big[ -\frac{\hbar ^2}{2m}\, 
\Big( \boldsymbol{\nabla} -i\bmatA -\frac{ie{\bf A}}{\hbar c}\Big) ^2
+W({\bf r})-gM_0\, \bsig \cdot {\bf n_0}\Big] \, \psi \, ,
\end{eqnarray}
where
\begin{eqnarray}
\label{4}
\bmatA ({\bf r})=iU\, (\boldsymbol{\nabla} U^{-1})	
\end{eqnarray}
is the gauge potential. Hamiltonian (S3) describes electrons moving in a homogeneous magnetization field $M_0{\bf n}_0$ under magnetic field ${\bf B}$ and the gauge potential $\bmatA ({\bf r})$.

In the case of magnetization along the ring with a constant value of $n_z$ (as shown in Figure 1a), the matrix of unitary transformation is
\begin{eqnarray}
\label{5}
U({\bf r})=e^{i\alpha \sigma _z} ,
\end{eqnarray}
where $\alpha ({\bf r})$ is the angle around circle. 
Then using Equation (S4) we obtain
\begin{eqnarray}
\label{6}
\bmatA ({\bf r})=\sigma _z\, \boldsymbol{\nabla}\, \alpha . 	
\end{eqnarray}
Using the angular coordinate $\alpha $ at the ring we present the one-particle Hamiltonian in form
\begin{eqnarray}
\label{7}
H=-\frac{\hbar ^2}{2mR^2}\, \Big( \frac{d}{d\alpha }-i\sigma _z-\frac{ieA_l}{\hbar c}\Big) ^2
-gM_0\, \bsig \cdot {\bf n}_0\, , 	
\end{eqnarray}
where $R$ is the ring radius and $A_l=BR/2$ is the longitudinal component (along the ring) of the vector potential ${\bf A}$. 
The eigenfunction of operator (S7) has the form $\psi (\alpha )=e^{is\alpha }\psi _s$, where $s$ is the integer quantum number. Correspondingly, the Schr\"odinger equation for spinor $\psi _s$ is 
\begin{eqnarray}
\label{8}
\Big[ \frac{\hbar ^2}{2mR^2}\, \Big( s-\sigma _z-\frac{eBR^2}{2\hbar c}\Big) ^2
-gM_0\, \bsig \cdot {\bf n}_0-\varepsilon \Big] \psi _s=0,  	
\end{eqnarray}
from which we obtain the energy of electron in $s$-state
\begin{eqnarray}
\label{9}
\varepsilon _s=\frac{\hbar ^2(\tilde{s}^2+1)}{2mR^2}\pm
\left[ \Big( \frac{\hbar ^2\tilde{s}}{mR^2}-gM_z\Big) ^2+g^2M_l^2\right] ^{1/2} ,	
\end{eqnarray}
where $\tilde{s}=s-\Phi /\Phi _0$, $\Phi =\pi R^2B$ is the magnetic flux through the ring and $\Phi _0=hc/e$ is the elementary flux.
It can be also presented as a dependence of $\varepsilon _s$ on the angle $\theta $ between vector ${\bf M}$ and axis $z$
\begin{eqnarray}
\label{10}
\varepsilon _s(\theta )=\frac{\hbar ^2(\tilde{s}^2+1)}{2mR^2}\pm
\left[ \Big( \frac{\hbar ^2\tilde{s}}{mR^2}-gM_0\cos \theta \Big) ^2+g^2M_0^2\sin ^2\theta \right] ^{1/2} .	
\end{eqnarray}
We can use the notations
$\varepsilon _0=\hbar ^2/mR^2,\; \xi =gM_0/\varepsilon _0$, and $t=\cos \theta $.
The parameter $\xi $ describes a ratio of the spin splitting to the size quantization splitting $\varepsilon _0$.   
Then we get
\begin{eqnarray}
\label{11}
\frac{\varepsilon _s}{\varepsilon _0}
=\frac{\tilde{s}^2+1}2 \pm\big( \tilde{s}^2-2\tilde{s}\xi t +\xi ^2\big) ^{1/2}.
\end{eqnarray}  
As we see, the energy of electron at the quantum level $s$ is depending on the orientation of magnetization of the ring. 
Figure 2b shows how the energy is depending on magnetic field at different orientations of magnetization. The dependence of $\varepsilon _s$ with $s=1$ on the parameter $t$ for different values of field is presented in Figure 2a. We see that when the magnetic field is small, the electron energy has a minimum at $t=0$ (for in-plane magnetization). But with the increasing field, after certain critical value $\Phi _c$ of the flux, the electron energy gets smaller if $t=1$. Thus, there appear an energy gain related to electron system, which makes it favorable the magnetization jump -- the magnetization changes its orientation from in-plane to out-of-plane along axis $z$. Besides, as the jump is associated with the critical flux $\Phi _c$, this imposes a relation between the magnetic field $B_c$ and the cell size $S$. For the parameters in Figure 2a, we have a critical point $B_c\simeq \Phi _0/S$. For example, if we have two different contours with $S_2=7S_1$ (it means that $S_1$ is one-cell area and $S_2$ has 7 cells) then $B_{c1}/B_{c2}=7$).

The obtained results can be interpreted in terms of the Berry phase of electron moving along the contour C within the ring. Indeed, the wave function of electron moving adiabatically in a non-homogeneous magnetization field ${\bf M}({\bf r})$ acquires the Berry phase $\gamma ^g_C=\int _C \mathcal{A}_l\, dl$, where $\mathcal{A} _l$ is the longitudinal component of gauge potential (S6). For the closed contour (ring) we find $\gamma ^{g}_C=\oint _C \mathcal{A}_l\, dl=2\pi $, which is one-half of the full surface of Berry sphere (the mapping space of vector field ${\bf n}({\bf r})$). This corresponds to the flux of topological (gauge) field $\bmatB =\boldsymbol\nabla \times \bmatA $ created by the monopole with topological charge $e_m=1$ in the center of Berry sphere.     

The magnetic field ${\bf B}$ generates an additional (Aharonov-Bohm) phase $\gamma ^m_C=\oint _C A_l\, dl$ at the same closed contour C, so that the total flux $\gamma _C=\gamma ^g_C+\gamma ^m_C$ through the one-half of Berry sphere is not $2\pi $ but is depending on ${\bf B}$. This can be viewed as a variation of monopole charge, which changes the total flux through one-half of the Berry sphere.

\section{Mechanism of anomalous Hall effect in a 2D electron system with magnetized honeycomb lattice}   

\begin{figure}
\centering
\includegraphics[width=\linewidth]{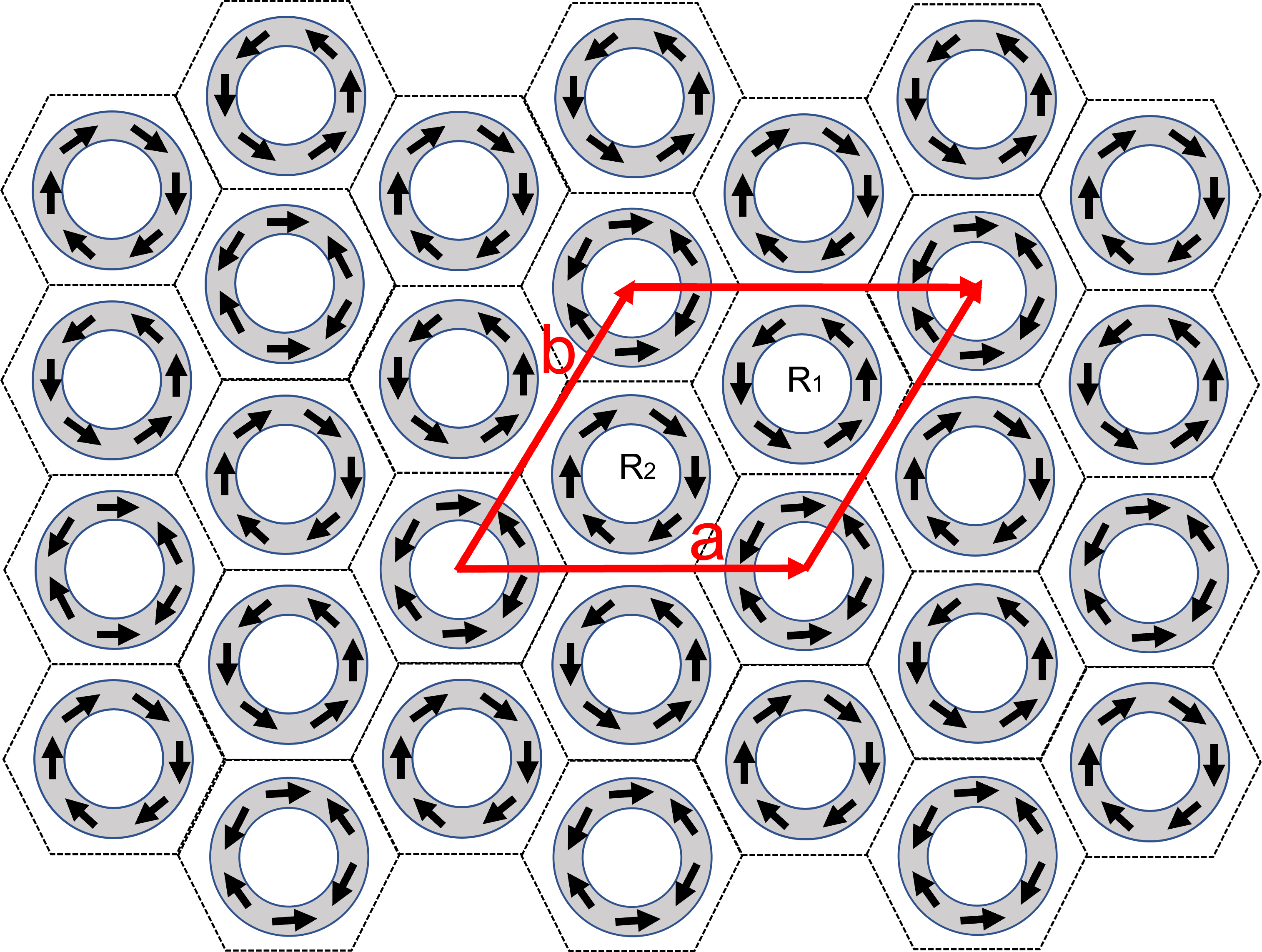}
\caption{The spin solid model of 2D electron gas (2DEG) with magnetized rings of alternating chiralities in a given magnetic unit cell.}
\label{fig:spin_solid_model}
\end{figure}

Here we propose a possible mechanism of anomalous Hall effect in a two-dimensional electron system on a magnetic network with the inhomogeneous in-plane magnetization. The key role of proposed effect is a certain chirality of the magnetic structure. For definiteness we consider the model with Hamiltonian
\begin{eqnarray}
\label{12}
H=\int d^2{\bf r}\; \psi ^\dag ({\bf r})\left[ 
-\frac{\hbar ^2(\nabla _x^2+\nabla _y^2)}{2m}+g\, \bsig \cdot {\bf M}({\bf r})\right] \psi ({\bf r})	
\end{eqnarray}
assuming that magnetization field ${\bf M}({\bf r})$ is forming a lattice of magnetic rings presented in Figure~\ref{fig:spin_solid_model},
\begin{eqnarray}
\label{eq:mag}
{\bf M}({\bf r})=\sum _i {\bf m}({\bf r}-{\bf R}_i^1)+{\bf m}({\bf r}-{\bf R}_i^2),
\end{eqnarray}
where the magnetization of a single ring 
\begin{eqnarray}
\label{14}
{\bf m}({\bf r})=\frac{\lambda _0}{r^2}\, (\hat{\bf z}\times {\bf r})
\end{eqnarray}
for $r_1<r<r_2$ and ${\bf m}({\bf r})=0$ otherwise. Vector ${\bf R}_i^1$ and ${\bf R}_i^2$ in Equation~(\ref{eq:mag}) determines the locations of the center of single rings of opposite chiralities in the $i$th unit cell. The Fourier transformation of Equation~(\ref{eq:mag}) is  
\begin{eqnarray}
\label{15}
{\bf M}({\bf q})=i\lambda _q \sum _i ( e^{-i{\bf q}\cdot {\bf R}_i^1} - e^{-i{\bf q}\cdot {\bf R}_i^2})\, (\hat{\bf z}\times {\bf n}_{\bf q}), 
\end{eqnarray}
where $\hat{\bf z}$  is the unit vector along axis $z$ perpendicular to 2D plane, ${\bf n}_{\bf q}$ is the unit vector along ${\bf q}$ and we denoted 
\begin{eqnarray}
\label{16}
\lambda _q=\frac{2\pi \lambda _0}{q} \int _{qr_1}^{qr_2} J_1(x)\, dx.
\end{eqnarray}
In Equation~(S16), $J_1(x)$ is the Bessel function. Using Equation (S16) we can present the magnetization profile in the following form
\begin{eqnarray}
\label{17}
{\bf M}({\bf r})=i\int \frac{d^2{\bf q}}{(2\pi )^2}\, \lambda _q
\sum _i (e^{i{\bf q}\cdot ({\bf r}-{\bf R}_i^1)} - e^{i{\bf q}\cdot ({\bf r}-{\bf R}_i^2)} )\, (\hat{\bf z}\times {\bf n}_{\bf q}) .
\end{eqnarray}
Correspondingly, the matrix element of perturbation related to magnetic lattice
\begin{eqnarray}
\label{18}
V_{\bf kk'}=\frac{ig\lambda _{{\bf k-k'}}}{\Omega _0}\sum _i (e^{-i({\bf k-k'})\cdot {\bf R}_i^1} - e^{-i({\bf k-k'})\cdot {\bf R}_i^2})\; 
\bsig \cdot (\hat{\bf z}\times {\bf n}_{\bf k-k'}), 
\end{eqnarray}
where $\Omega _0$ is the sample area.

\begin{figure}
\centering
\includegraphics[width=\linewidth]{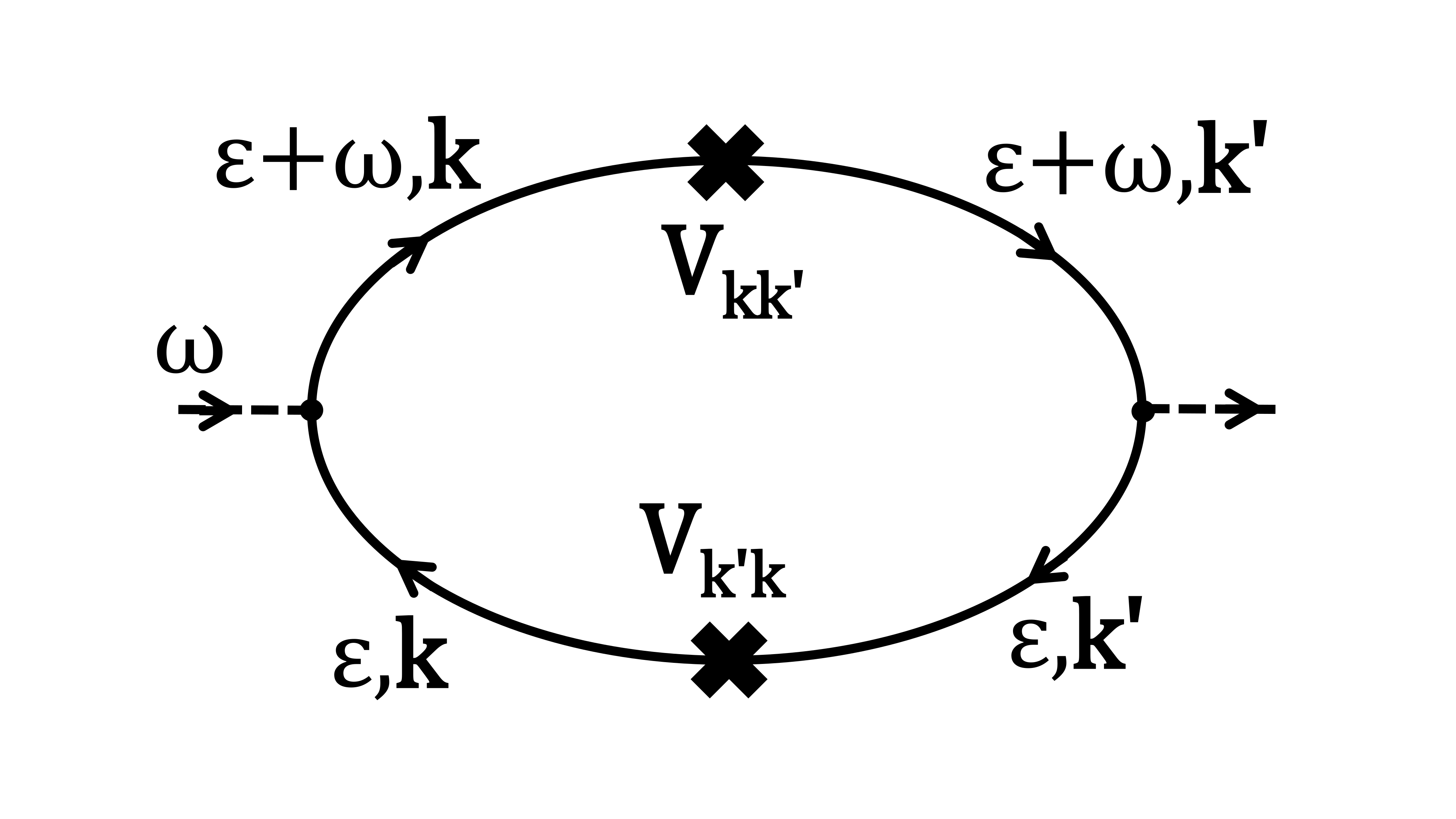}
\caption{Kubo diagram for the conductivity.}
\label{fig:diagram}
\end{figure}

\begin{figure}
\centering
\includegraphics[width=\linewidth]{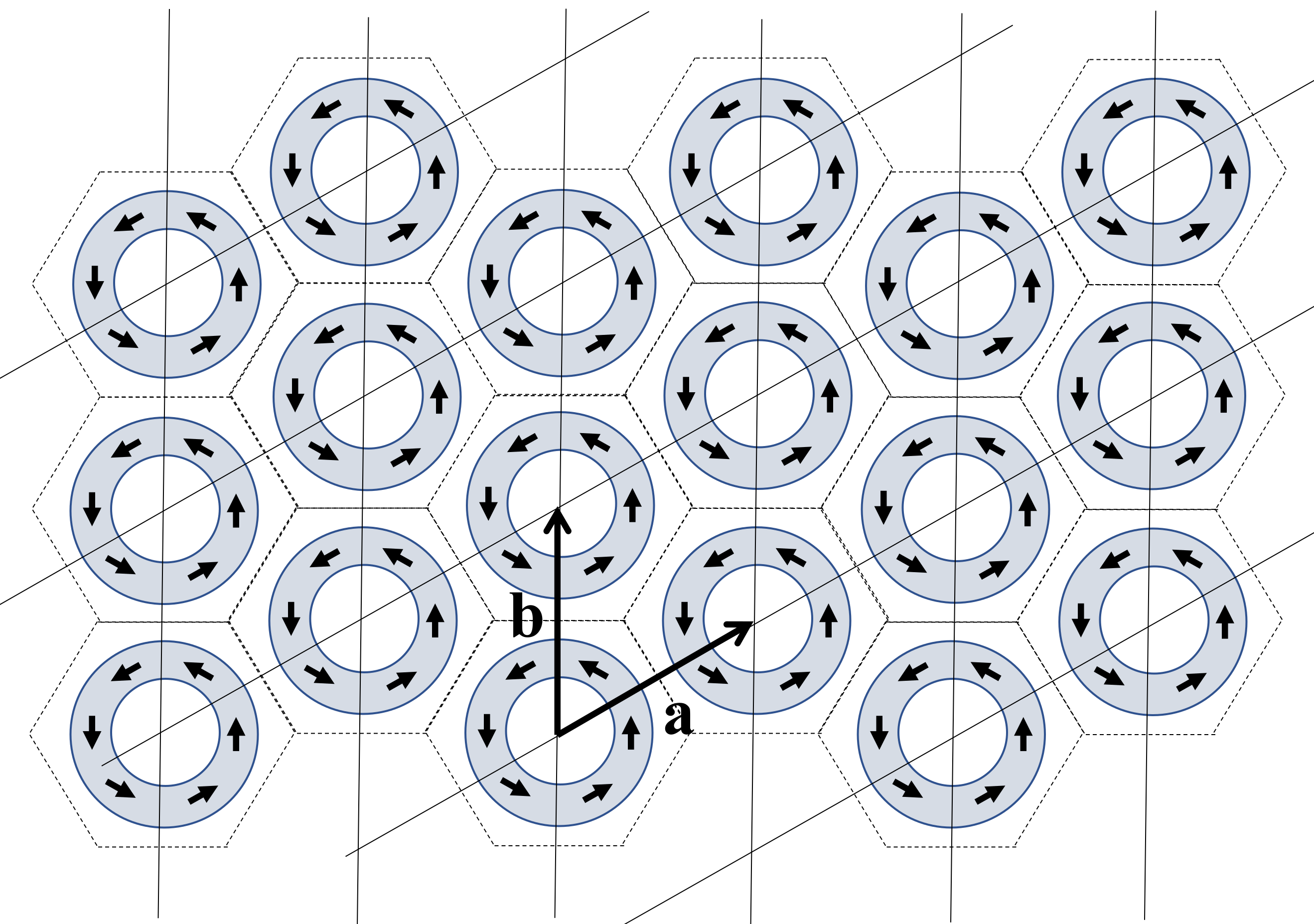}
\caption{The model of 2DEG with a lattice of magnetized rings of the same chirality. Each magnetic unit cell contains one magnetized ring.}
\label{fig:simple_model}
\end{figure}

\begin{figure}
\centering
\includegraphics[width=\linewidth]{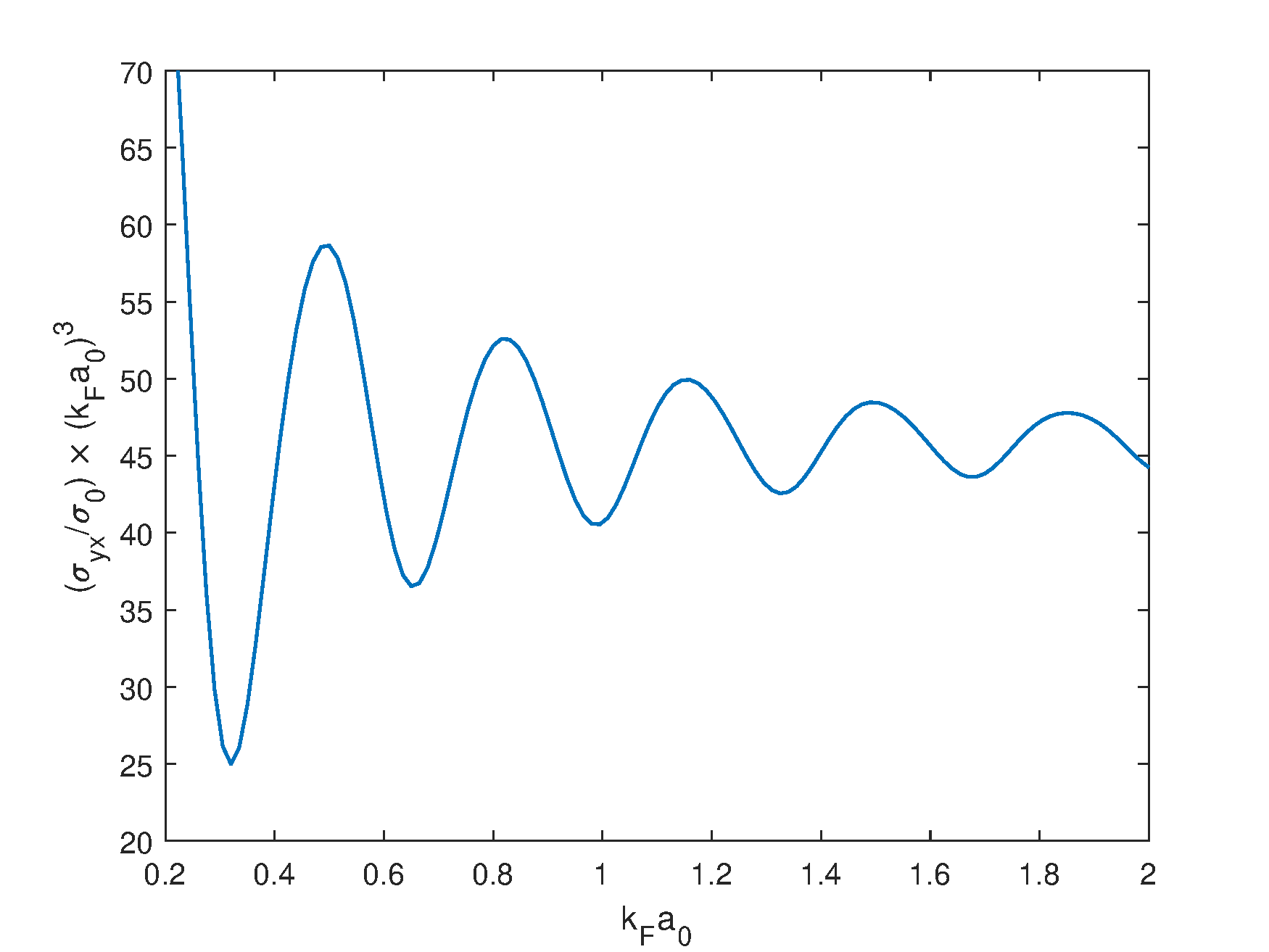}
\caption{First principle calculation of the transverse conductivity based on the 2DEG model presented in Figure~\ref{fig:simple_model}.}
\label{fig:simple_model_conduct}
\end{figure}

Now we calculate the current along axis $y$ assuming the electric field applied along axis $x$. 
Using the Kubo method we present the transverse current 
\begin{eqnarray}
\label{19}
j_{y\omega }=-\frac{ie^2\hbar ^2A_{x\omega }}{m^2c}\, {\rm Tr} 
\int d^2{\bf r}_1\int \frac{d\varepsilon }{2\pi }\; 
\nabla _y\; \tilde{G}({\bf r,r}_1;\varepsilon +\hbar \omega )\; \nabla _x\; 
\tilde{G}({\bf r}_1,{\bf r'};\varepsilon )\Big| _{{\bf r'}={\bf r}} ,
\end{eqnarray}
where $G({\bf r,r'}; \varepsilon )$ is the Green's function of Hamiltonian (S12). One can calculate the transverse current taking the magnetic lattice in second-order approximation of perturbation theory as presented in diagram of Figure~\ref{fig:diagram}. \cite{note1} Substituting $A_{x\omega }=cE_{x\omega }/i\omega $ and taking the limit $\omega \to 0$ we obtain 
\begin{eqnarray}
\label{20}
j_y=\frac{e^2\hbar ^2E_x\Omega _0}{m^2\omega }\, \lim _{\omega \to 0} {\rm Tr} 
\int \frac{d^2{\bf k}}{(2\pi )^2} \frac{d^2{\bf k'}}{(2\pi )^2} \frac{d\varepsilon }{2\pi }\; 
k_y\, G_{0k}(\varepsilon +\hbar \omega )\, V_{\bf kk'}\, G_{0k'}(\varepsilon +\hbar \omega )\, k'_x\,
G_{0k'}(\varepsilon )\, V_{\bf k'k}\, G_{0k}(\varepsilon ) ,
\end{eqnarray}
 where 
\begin{eqnarray}
\label{21}
G_{0k}(\varepsilon )=\big( \varepsilon -\varepsilon _k+\mu +i\Gamma \, {\rm sgn}\, \varepsilon \big) ^{-1} 
\end{eqnarray}
is the Green's function of free electron, $\Gamma =\hbar /2\tau $ is the relaxation rate, $\tau $ the electron relaxation time, $\varepsilon _k=\hbar ^2k^2/2m$, and $\mu $ is the chemical potential. The integral over $\varepsilon $ is not zero if the Green's function poles are in different halfplanes of complex $\varepsilon $. Then we get
\begin{eqnarray}
\label{22}
j_y=\frac{e^2\hbar ^3E_x\Omega _0}{2\pi m^2}\, {\rm Tr}
\int \frac{d^2{\bf k}}{(2\pi )^2} \frac{d^2{\bf k'}}{(2\pi )^2}\,
k_y\, G^R_{0k}\, V_{\bf kk'}\, G^R_{0k'}\, k'_x\,
G^A_{0k'}\, V_{\bf k'k}\, G^A_{0k}\, ,
\end{eqnarray}
where $G^{R,A}_k$ are the retarded and advanced Green's functions at $\varepsilon =0$. Using Equation (S18) we find 
\begin{equation}
\label{eq:conduct}
\begin{split}
\sigma _{yx}=\frac{e^2\hbar ^3g^2N}{\pi m^2\Omega _0}
&\int \frac{d^2{\bf k}}{(2\pi )^2}\, 
\int \frac{d^2{\bf k'}}{(2\pi )^2}\,
\lambda ^2_{\bf k-k'}\,
k_y\, G^R_k\, G^A_k\,
k_x'\, G^R_{k'}\, G^A_{k'}\,\\
&\sum _{nn'}(e^{-i({\bf k}-{\bf k'})\cdot (n {\bf a} + n'{\bf b})}(2-e^{-i({\bf k}-{\bf k'})\cdot \boldsymbol{\delta}}-e^{-i({\bf k'}-{\bf k})\cdot \boldsymbol{\delta}} ),
\end{split}
\end{equation}
where $N$ is the number of cells. Basis vectors of the honeycomb lattice presented in Figure~\ref{fig:spin_solid_model} are
\begin{eqnarray}
\label{eq:lattice_vectors}
{\bf a}=a_0\; (1,\, 0),\hskip0.3cm
{\bf b}=\frac{a_0}{2}\, (1,\, \sqrt{3}), \hskip0.3cm \boldsymbol{\delta}=\frac{a_0}{2}(1,\, \frac{\sqrt{3}}{3}),
\end{eqnarray}
where $a_0$ is the lattice constant of magnetic structure and $\boldsymbol{\delta} = {\bf R}_i^1 - {\bf R}_j^2$, $n$ and $n'$ are integers. Then we obtain the transverse conductivity consists of three terms
\begin{eqnarray}
\label{25}
\sigma _{yx} = \sigma_{yx}^1 - \sigma_{yx}^2 -\sigma_{yx}^3,
\end{eqnarray}
where
\begin{eqnarray}
\label{26}
\sigma_{yz}^1=\frac{e^2\hbar ^3g^2}{\pi m^2a_0^2}\sum _{nn'}
\int \frac{d^2{\bf k}}{(2\pi )^2}\,
2e^{-\frac{ika_0}2 ((2n+n')\cos \varphi +\sqrt{3}n'\sin \varphi )}\; 
k\sin \varphi \; G^R_k\, G^A_k
\nonumber \\ \times
\int \frac{d^2{\bf k'}}{(2\pi )^2}\, \lambda ^2_{\bf k-k'}\,
e^{\frac{ika_0}2 ((2n+n')\cos \psi +\sqrt{3}n'\sin \psi)}\; 
k'\cos \psi \; G^R_{k'}\, G^A_{k'}\, .
\end{eqnarray}
Then after integrating over $k$ we get
\begin{equation}
\label{27}
\begin{split}
\sigma _{yx}^1
\simeq 2\frac{e^2g^2\lambda ^2k_F^2\tau ^2}{4\pi ^3\hbar ^3a_0^2}\sum _{nn'}&
\int _0^{2\pi } d\varphi \; e^{-\frac{ik_Fa_0}2 ((2n+n')\cos \varphi +\sqrt{3}'n\sin \varphi)} \sin \varphi \\
&\int _0^{2\pi } d\psi \; e^{\frac{ik_Fa_0}2 ((2n+n')\cos \psi +\sqrt{3}n'\sin \psi)} \cos \psi ,
\end{split}
\end{equation}
where we introduced the mean square of $\lambda ^2_{\bf k-k'}$ for $k,k'=k_F$
\begin{eqnarray}
\label{28}
\lambda ^2=\frac1{\pi }\int _0^\pi d\phi \, \lambda ^2_{{\bf k}_F-{\bf k}'_F}
=\frac1{\pi }\int _0^\pi d\phi \; \lambda ^2_{2k_F|\sin (\phi /2)|}.
\end{eqnarray}
Substituting Equation~(\ref{eq:lattice_vectors}) into~(\ref{eq:conduct}) results in similar expressions for $\sigma_{yx}^2$ and $\sigma_{yx}^3$.
The result of calculation is presented as a dependence of 
$(\sigma _{yx}/\sigma _0)$ on $k_Fa_0$. We denoted
$\sigma _0=e^2g^2\lambda ^2\tau ^2/4\pi ^3\hbar ^3a_0^4$.
The calculation is performed for a finite structure with $n$ and $n'$ running from $-10$ to $10$. As we see from Figure 5a, $(\sigma _{yx}/\sigma _0)$ oscillates as a function of $k_Fa_0$.
Thus, it is expected that any external perturbation, such as magnetic field application, that alters the Fermi level of the system or the magnetic unit cell size will lead to oscillatory transverse current. We have performed a similar calculations on another magnetization distribution, where each magnetic unit cell contains one magnetic ring of the same chirality, presented in Figure~\ref{fig:simple_model}. As we see from Figure~\ref{fig:simple_model_conduct}, $(\sigma _{yx}/\sigma _0) \times (k_Fa_0)^3$ is an oscillating function decreasing with $a_0$ as $1/a_0^3$.  
Thus, the transverse current can be rather strong in structures with small magnetic cells.   

The physical mechanism of anomalous Hall effect can be also understood from Equation~(S3). The gauge field $\bmatA $ comes to this equation like the vector potential of electromagnetic field ${\bf A}$. It affects the energy spectrum and electron wavefunctions of the system. On the other hand, nonzero circulation of the gauge field $\bmatA $ along the ring is equivalent to the flux of Berry curvature $\bmatB =\boldsymbol{\nabla} \times \bmatA $ penetrating the ring. Thus, the gauge field $\bmatB $ is acting like external magnetic field,  inducing the Hall current.

\section{Persistent currents}

Here we calculate the equilibrium persistent current in a single magnetic ring. We assume that in the equilibrium state without external field, the magnetic moments are in-plane along the ring ($M_z=0$). The Hamiltonian of electrons in the ring of radius $R$ is
\begin{eqnarray}
\label{29}
H=Rd_0\int d\alpha \; \psi ^\dag (\alpha )
\left[ -\frac{\hbar ^2}{2mR^2}\, \frac{d^2}{d\alpha ^2}
+gM\left( \begin{array}{cc} 0 & -ie^{-i\alpha } \\ ie^{i\alpha } & 0 \end{array}\right)  \right] \psi (\alpha ),
\end{eqnarray}
where $d_0$ is the width of ring. The eigenfunctions and eigenvectors can be found from the Schr\"odinger equation 
\begin{eqnarray}
\label{30}
\left( \begin{array}{cc} 
-\frac{\hbar ^2}{2mR^2}\, \frac{d^2}{d\alpha ^2} -\varepsilon & 
-igMe^{-i\alpha } \\ igMe^{i\alpha } & 
-\frac{\hbar ^2}{2mR^2}\, \frac{d^2}{d\alpha ^2}-\varepsilon 
\end{array}\right)  
\left( \begin{array}{c} c_1\, e^{in\alpha } \\ c_2\, e^{i(n+1)\alpha }\end{array}\right) =0,
\end{eqnarray}
from which we find eigenenergies
\begin{eqnarray}
\label{31}
\varepsilon _n=\frac{\hbar ^2}{4mR^2}\, \big[ n^2+(n+1)^2\big]
\pm \Big( \frac{\hbar ^4}{16m^2R^4}\, \big[ n^2-(n+1)^2\big] ^2+g^2M^2\Big) ^{1/2} 
\end{eqnarray}
and the relation between coefficients $c_1$ and $c_2$
\begin{eqnarray}
\label{32}
c_2=-\frac{i}{gM}\, \Big( \frac{\hbar ^2n^2}{2mR^2}-\varepsilon _n\Big) \, c_1.
\end{eqnarray}
Then using the normalization condition, $|c_1|^2+|c_2|^2=1$, we find
\begin{eqnarray}
\label{33}
|c_1|^2=\frac1{1+\frac1{g^2M^2} \Big( \frac{\hbar ^2n^2}{2mR^2}-\varepsilon _n\Big) ^2}\, ,
\\
|c_2|^2=\frac{\frac1{g^2M^2} \Big( \frac{\hbar ^2n^2}{2mR^2}-\varepsilon _n\Big) ^2}
{1+\frac1{g^2M^2} \Big( \frac{\hbar ^2n^2}{2mR^2}-\varepsilon _n\Big) ^2}\, .
\end{eqnarray}
Using the current operator, corresponding to electric current along the ring
\begin{eqnarray}
\label{34}
\hat{j}_\alpha =-\frac{ie\hbar \nabla _\alpha }{mR} 
\end{eqnarray}
and the eigenfunctions of Hamiltonian (S29) we can calculate the expectation value of $\hat{j}_\alpha $
\begin{eqnarray}
\label{35}
j_\alpha =-\frac{ie\hbar }{mR}
\sum _n f(\varepsilon _n)\, \big( \psi ^\dag _n\, \nabla _\alpha \, \psi _n\big)
=\frac{e\hbar }{mR}\sum _n\, \big[ n|c_1|^2+(n+1)|c_2|^2\big] \, f(\varepsilon _n),    	
\end{eqnarray}
where $\psi _n^T=\big( c_1\, e^{in\alpha },\, c_2\, e^{i(n+1)\alpha }\big) $ and $f(\varepsilon _n)$ is the Fermi-Dirac function. The dependence of current $j_\alpha $ (in units $e\hbar /mR$) on the chemical potential $\mu $ for $gM=0.01$~eV is presented in Figure 5b.  

The Berry phase at the contour along the ring
\begin{eqnarray}
\label{36}
\phi _n=\int _0^{2\pi }d\alpha \, A_n(\alpha ),	
\end{eqnarray}
where $A_n(\alpha )$ is the gauge potential
\begin{eqnarray}
\label{37}
A_n(\alpha )=-i \psi ^\dag _n\, \nabla _\alpha \, \psi _n\, .	
\end{eqnarray} 
Using Equation~(S35) we can present the equilibrium current by the Berry phase
\begin{eqnarray}
\label{38}
j_\alpha =\frac{e\hbar }{2\pi mR}\sum _n f(\varepsilon _n)\, \phi _n\, . 	
\end{eqnarray}
The existence of equilibrium current in the nano ring is related to inhomogeneous magnetization. Indeed, in the noncollinear magnetic state one appears the spin torque acting on magnetic moments. The torque transfer can be viewed as the spin current of propagating electrons, which, in its turn, is accompanied by the charge current due to the imbalance of electrons with different spin polarization.

\clearpage

\clearpage
\begin{figure}
\centering
\includegraphics[width=\linewidth]{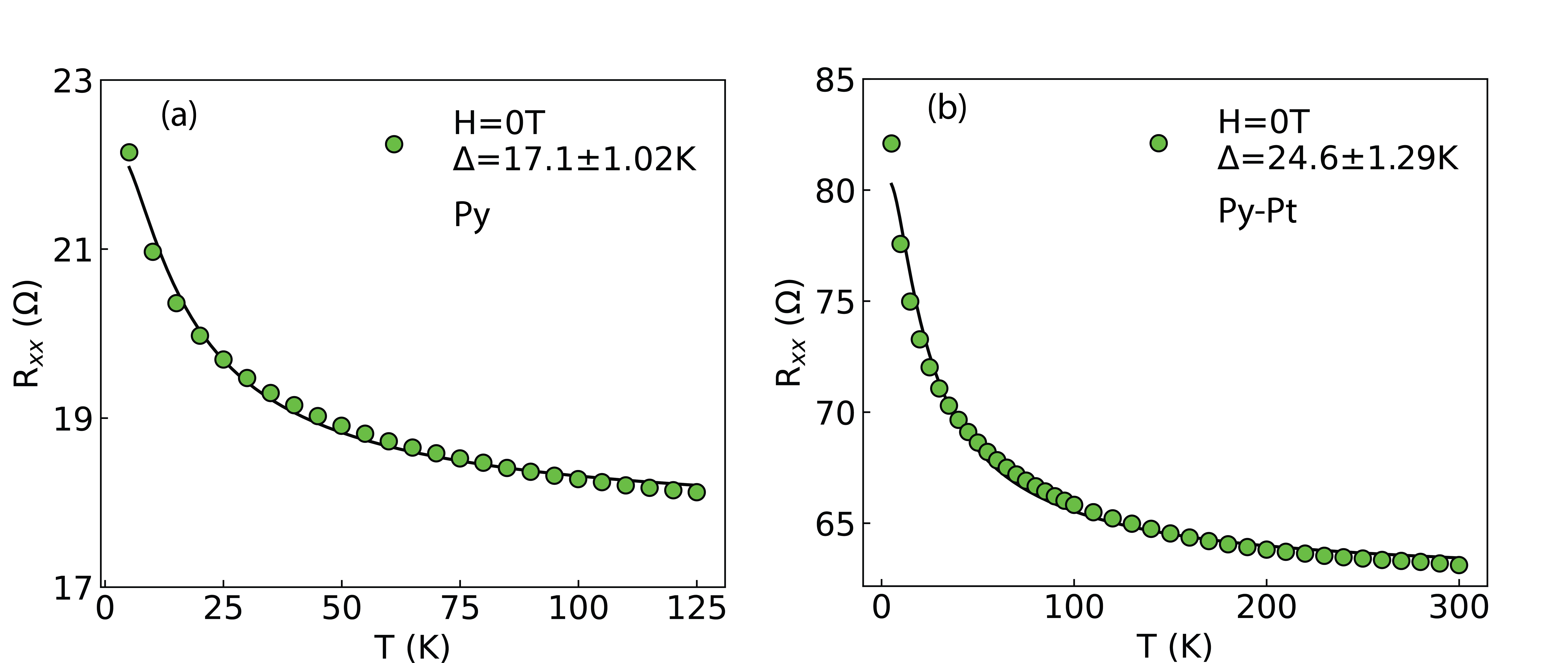}
\caption{Semiconducting-type linear resistance fitted with Arrhenius law. The activation energies are found to be around 17 K and 25 K in temperature unit for (a) Py honeycomb and (b) Py-Pt honeycomb, respectively.}
\label{fig:resistivity_fitting}
\end{figure}